\documentclass[12pt]{article}
             \usepackage{amsfonts}
             \usepackage{amssymb}
\newtheorem{thm}{Theorem}
\newtheorem{cor}[thm]{Corollary}

\newtheorem{defn}[thm]{Definition}

\def\qed{{\bf QED}}

\def\tr{\hbox{Tr}}

\def\be{\begin{eqnarray}}
\def\ee{\end{eqnarray}}
\def\bee{\begin{eqnarray*}}
\def\eee{\end{eqnarray*}}

\def\ts{\textstyle}
\def\bra{\langle}
\def\ket{\rangle}
\def\kb{ \ket \bra }

\def\rt2{\ts \frac{1}{\sqrt{2}} }

\def\raw{\rightarrow}

\def\ot{\otimes}        
     \def\wh{\widehat}

    \def\ts{\textstyle}

\title{Multiplicativity properties of entrywise positive maps}

    \author{Christopher King and Michael Nathanson \\
    Department of Mathematics\\
Northeastern University,  Boston MA 02115 \\
\and Mary Beth Ruskai
      \\ Department of Mathematics \\
Tufts University,
     Medford, MA 02155 \\ }

\begin{document}

\maketitle

  \begin{abstract}
Multiplicativity of certain maximal $p
\rightarrow q$ norms of a tensor product of linear maps on
matrix algebras is proved in situations in which  the condition of
complete positivity (CP) is either  augmented by,
or replaced by, the requirement  that the entries of a
matrix representative of  the map are non-negative (EP).
In particular, for integer $t$,  multiplicativity holds for the maximal
$2 \rightarrow 2t$ norm of a product of two maps,
whenever one of the pair is EP; for the maximal $1 \raw t$ norm
for pairs of CP maps when one of them is also EP; and for the maximal
$1 \raw 2t$ norm for the product of an EP and a 2-positive map.
Similar results are shown in the infinite-dimensional setting of
convolution operators on $L^2({\Bbb R})$,
with the pointwise positivity of an integral kernel replacing entrywise
positivity of a matrix. These results apply in particular to Gaussian bosonic
channels.
\end{abstract}

\section{Introduction}
The additivity conjecture for
minimal output entropy of product channels remains a challenging open
problem in quantum information
theory \cite{Shor2}. In this paper we study a class of completely
positive (CP) maps for which the related
question of multiplicativity of maximal output purity \cite{AHW} can
be demonstrated for
    {\em integer values} of the parameter $p$.
The multiplicativity property follows from the existence of a basis
in which the map satisfies a condition we call entrywise positive (EP),
so that H\"older's inequality can  be applied in a useful
way. Several classes of maps satisfying
the EP property are presented.

\section{Statement of results}
Throughout we will denote by $M_{n}$ the vector space of
complex-valued $n \times n$ matrices. The Schatten norm of $A \in
M_n$ is defined for $p \geq 1$ by
\be
|| A ||_p = \bigg( \tr \, |A|^p \, \bigg)^{1/p}
\ee
For a linear map $K: M_n \rightarrow M_m$ and $p,q \geq 1$ we define
the family of norms
\be\label{def:p,q-norm}
|| K ||_{p \rightarrow q} = \sup \,\, \Bigg\{ { || K(A) ||_q \over ||
A ||_p } \,\, : \,\,
A \neq 0 \Bigg\} ~.
\ee
One can also consider such norms when $A$ is restricted to the real vector
space of self-adjoint matrices.   We will not do this here, except for the case
$p = 1$ which we denote by   ${\nu}_q (K)$.
As noted in \cite{AH, KR2}, this is equivalent to
\be\label{def:nu}
{\nu}_q (K) = \sup \,\, \Bigg\{ { || K(A) ||_q \over \tr \, A } \,\, : \,\,
A \geq 0, A \neq 0 \Bigg\}.
\ee

\begin{defn}
A linear  map $\Phi: M_n \mapsto M_m$ is called {\em entrywise positive}
{\rm (EP)}  if all entries of $\Phi$ are nonnegative with respect to
some pair of orthonormal bases $\{ | e_j \ket \}$ and $\{ | f_k \ket \}$ for
    ${\Bbb C}_n$ and ${\Bbb C}_m$, respectively.  That is,
\be\label{def:EP}
\tr \,\, | f_k  \kb f_{\ell} | \, \Phi \big( | e_i \kb e_j | \big) =
\bra f_{\ell}  | \, \Phi \big( | e_i \kb e_j | \big) \, | f_k \ket  \geq 0
\ee
for all $i,j,k, \ell$.
\end{defn}

    Recall that the matrix representative of a linear operator
   using  orthonormal bases
    $\{ | e_j \ket \}$ and $\{ | f_k \ket \}$ for its domain and range
    is $a_{jk} = \bra f_k, A e_j \ket$.   Thus, for each fixed $i,j$
    the expression (\ref{def:EP}) gives the $\ell, k$ entry of the
$m \times m$  matrix representative for the operator
$\Phi( | e_i \kb  e_j  | )$, i.e, the $\ell, k$ entry of the
$i, j$ block in the $mn \times mn$ Choi-Jamiolkowski matrix   (or 
state representative of $\Phi$).
Alternatively, one can regard (\ref{def:EP}) as describing the
$m^2 \times n^2$ matrix representative of $\Phi$ using  input
and output bases with the form $ E_{ij} = |e_i \kb e_j | $ and
$F_{k \ell} = |f_k \kb f_{\ell} |$, respectively.    The condition that the
$m^2 n^2$ numbers given by (\ref{def:EP}) are non-negative is independent
of whether or not they are arranged in any particular matrix form.

   The condition (\ref{def:EP})  can be restated as follows.
   Let $\Gamma_U$ denote the map which acts by
conjugation with $U$, that is
$ \Gamma_U (Q) = U Q U^{*}$.
   Then $\Phi$ is EP if there are unitary matrices $U \in M_n$ and
   $V \in M_m$ such that the operator
   $\Gamma_V \circ \Phi \circ  \Gamma_{U}$
   satisfies  (\ref{def:EP}) in the standard bases for ${\Bbb C}_n$ and
${\Bbb C}_m$.

   \medskip
The entrywise positivity property also arises
for integral 
operators on function spaces,
where it is expressed as pointwise 
positivity of
the integral kernel. In this context,
multiplicativity of the $p \rightarrow q$ norm
defined in (\ref{def:p,q-norm}) was proved for integral operators
for all $1 \leq p \leq q$ \cite{Beck, Lieb}. Using the 
additional
assumption of pointwise positivity, Lieb extended this 
result
to all $1 \leq p, q$
 (see Theorem 3.2 in \cite{Lieb}.    Although the proofs in \cite{Beck, Lieb} are given
for the tensor product of a kernel with itself, the argument extends
to different kernels.)   Restricting to finite dimensions yields
multiplicativity for
linear maps acting on the subalgebra of diagonal
matrices, where again the result holds in general for $1 \leq p \leq q$,
and under the EP assumption for all $1 \leq p, q$.

\medskip
Less is known about multiplicativity for maps on the full
matrix algebra. In the following theorems we use the EP
property to demonstrate
this property in several cases. The first result applies to
linear maps on matrix
algebras without the additional assumption of complete positivity.
It asserts multiplicativity of the maximal $2 \raw 2t$ norm 
for integer $t$ whenever one of the
maps is EP, and provides an upper bound on the $p \raw 2t$ norm
for $1 \leq p \leq 2$.

\begin{thm}\label{thm1}
Let $K$ and $L$ be linear maps on $M_n$ and $M_m$ respectively, and
suppose that
$K$ is {\rm EP}. Then for all $1 \leq p \leq 2$, and all
integers $t $,
\be\label{thm1:eqn1}
|| K \ot L ||_{p \rightarrow 2 t} \leq || K ||_{2 \rightarrow 2 t}
\,\, || L ||_{p \rightarrow 2 t}
\ee
with equality when $p=2$.
\end{thm}

The next result uses the assumption of complete positivity to deduce a
multiplicativity result for  (\ref{def:nu}), which is the case of
interest in quantum
information theory.

\begin{thm}\label{thm2}
Let $\Phi$ and $\Omega$ be  CP maps on $M_n$ and
$M_m$ respectively,
and assume that $\Phi$ is also  {\rm EP}. Then for all integers $t$,
\be\label{thm2:eqn1}
\nu_{t}( \Phi \ot \Omega ) = \nu_{t}( \Phi ) \,\, \nu_t (\Omega).
\ee
\end{thm}
  When $t$ is an even integer the hypothesis of Theorem~\ref{thm2}
  can be  weakened; the requirement that $\Phi$ be CP is not
  necessary and $\Omega$ need only be 2-positive.
  \begin{thm}\label{thm4}
Let $\Phi$ be an EP linear map and
let $\Omega$ be   a 2-positive map on
$M_n$ and $M_m$ respectively.
Then for all integers $t$,
\be\label{thm4:eqn1}
\nu_{2t}( \Phi \ot \Omega ) = \nu_{2t}( \Phi ) \,\, \nu_{2t} (\Omega).
\ee
\end{thm}

\medskip
It remains an open question whether the equality (\ref{thm2:eqn1}) holds
for other values of $t$, in particular for the range $1 \leq t \leq 2$.
It can be shown that (\ref{thm2:eqn1}) is true at $t=2$
under the weaker condition that $\widehat{\Phi} \circ \Phi$ is EP  \cite{KR2},
where $\widehat{\Phi}$ is the adjoint with respect to the
Hilbert-Schmidt inner product
(note however that (\ref{thm2:eqn1}) does {\em not} hold for general 
integer $t$
under this weaker condition, as  demonstrated  by the
well-known example of Holevo-Werner maps
\cite{WH}).

\medskip
Our last results concern the one-particle Hilbert space $L^{2}({\Bbb R})$,
where  states are
represented by kernels $K(x,y)$ satisfying
$K(x,y) = \overline{K(y,x)}$,
\be
\int \int \overline{\psi(x)} K(x,y) \psi(y) dx dy \geq 0
\ee
for all $\psi \in L^{2}({\Bbb R})$, and
\be
\int K(x,x) d x = 1
\ee
In this setting, a linear map $\Phi$ is a convolution operator:
\be\label{def:inf}
\Phi \,\, :\,\,  K(x, y) \rightarrow \int \int \,  G(x,y; u,v) \, 
K(u, v) \, d u \, d v
\ee
The analog of the entrywise positive (EP) property in the finite-dimensional case
is pointwise positivity of the kernel $G$, that is
\be\label{G-pos}
G(x,y; u,v) \geq 0
\ee
for all $u,v,x,y \in {\bf R}$.

For integer $t$, the Schatten norm of a state is defined by
\be\label{def:t-norm-inf}
|| \rho ||_t =
(\tr {\rho}^t )^{1/t} = \bigg( \int \dots \int K(x_1,x_2) K(x_2,x_3)
\dots K(x_t,x_1) d x_1 \, \dots \, d x_t \bigg)^{1/t}
\ee
When $\rho$ is positive and trace class, (\ref{def:t-norm-inf}) is well-defined
for all integer $t$, and hence (\ref{def:nu}) extends to this case also.

\medskip
\begin{thm}\label{CP-EP-inf}
Let $\Phi$ be a completely positive map
defined as in (\ref{def:inf}), with kernel $G$ satisfying the
positivity condition (\ref{G-pos}). Let $\Omega$ be any other
CP map on $L^2({\Bbb R})$. Then for integer $t$,
\be\label{CP-EP-inf1}
\nu_{t}(\Phi \ot \Omega) = \nu_{t}(\Phi) \,\, \nu_{t}(\Omega)
\ee
\end{thm}

\medskip
As an application of the previous result, recall the definition of a 
bosonic channel
\cite{HW1, GGLMS}:
\be\label{def:bosonic}
N(\rho) = \int \, P(z) \, D(z) \, \rho D(z)^{\dagger} \, d z
\ee
Here $D(z)$ is the unitary displacement operator for a coherent state, 
which acts on
$L^{2}({\Bbb R})$ according to
\be
(D(z) \psi) (x) = (D(\alpha, \beta) \psi) (x) = e^{i \alpha x} \, 
\psi(x - \beta)
\ee
The function $P(z)=P(\alpha, \beta)$ is a probability density 
function on ${\Bbb R}^2$,
so  (\ref{def:bosonic}) defines a unital trace-preserving CP channel
on states over $L^{2}({\Bbb R})$.
In the main case of interest for applications $P(z)$ is 
a Gaussian density \cite{GGLMS,HW1},  and some multiplicativity results have been
proved under this assumption \cite{GL2,SEW}.
As our next result shows,  Theorem \ref{CP-EP-inf} can be applied to
bosonic channels of the form (\ref{def:bosonic}) in the case where
$P$ satisfies two positivity conditions:
\be\label{P-pos}
P(\alpha, \beta) \geq 0, \quad
\int e^{i \alpha x} \, P(\alpha, \beta)\, d \alpha \geq 0
\quad \mbox{for all} \,\, x, \beta
\ee
In particular note that (\ref{P-pos}) holds for any Gaussian density.

\medskip
\begin{thm}\label{bos-ch}
Let $N$ be a map of the form (\ref{def:bosonic}) where $P(z)$ satisfies
(\ref{P-pos}), and let $\Omega$ be any CP map on $L^2({\Bbb R})$.
Then for integer $t$,
\be\label{bos-ch-1}
\nu_{t}(\Phi \ot \Omega) = \nu_{t}(\Phi) \,\, \nu_{t}(\Omega)
\ee
\end{thm}

\section{Examples of CP-EP maps}

   \subsection{Quantum-Classical maps}
   A map $\Phi: M_n \mapsto M_m$ takes a quantum system to a classical
one if its range is contained in the subset of diagonal matrices.  In this
case, the map is EP if and only if it is CP.

\subsection{Qubit maps}
We use the diagonal representation of qubit maps introduced in
\cite{KR1} and used, e.g.,  in \cite{RSW}.
In this representation a qubit map $\Phi$ acts as follows:
\be  \label{qubit.canon}
\Phi \Big(I + \sum w_k \sigma_k \Big) = I + \sum (\lambda_k w_k + t_k) \sigma_k
\ee
where $\sigma_k$ are the Pauli matrices.
The Choi matrix of $\Phi$ in this representation is
\be
\begin{pmatrix}{\Phi(E_{11} )& \Phi(E_{12}) \cr
\Phi(E_{21}) & \Phi(E_{22})}
\end{pmatrix}
    = {1 \over 2} \,\,
\begin{pmatrix} {
1 + \lambda_3 + t_3 & t_1  - i  t_2 & 0 & \lambda_1 + \lambda_2 \cr
t_1  + i  t_2 & 1 - \lambda_3 - t_3 & \lambda_1 - \lambda_2 & 0 \cr
0 & \lambda_1 - \lambda_2 & 1 - \lambda_3 + t_3 & t_1 - i t_2 \cr
\lambda_1 + \lambda_2 & 0 & t_1 + i t_2 & 1 + \lambda_3 - t_3}
\end{pmatrix} \nonumber
\ee

The CP condition puts some constraints
on the six parameters $\{\lambda_k , t_k \}$, and these are fully
explored in \cite{RSW}.
By changing bases if necessary in the domain and range of $\Phi$,
(i.e., usng $\Gamma_U \circ   \Phi \circ \Gamma_V $  as discussed in \cite{King1,KR1}) 
it can be assumed that the
following conditions are satisfied:
\be
\lambda_1 \geq | \lambda_2 | , \quad
t_1 \geq 0
\ee
The EP condition is satisfied if
\be  \label{qt20}
\lambda_1 \geq | \lambda_2 | , \quad
t_1 \geq 0, \quad t_2 = 0
\ee
Hence the only additional restriction coming from the EP condition is 
$t_2 = 0$.
The index `2' here has a geometric meaning as it labels one of the
two smaller axes of the
image ellipsoid.  Theorem~\ref{thm2} then implies the following.
\begin{cor}\label{cor.qubit}
Let $\Phi$ be a qubit channel, and suppose that $t_2 = 0$ in the
diagonal representation,
where $\lambda_2$ describes one of the two smaller axes of the image
ellipsoid.  (If any two axes
have equal length, there is no restriction.). Then $\nu_{t}( \Phi \ot 
\Omega ) = \nu_{t}( \Phi ) \,\nu_t (\Omega)$ for any CP map $\Omega$, 
for all integer $t$.
\end{cor}
The methods of \cite{Ki5} can be used to extend this multiplicativity
result to all non-integer values
$p \geq 2$ for the same class of qubit channels. Unfortunately these
methods do not
apply for values $1 < p < 2$, and for this interval the
multiplicativity question for qubit maps is open
except for unital channels \cite{Ki3}.

\subsection{Depolarizing channels and generalizations}
The $d$-dimensional depolarizing channel is the map
\be\label{def:dep}
\rho \rightarrow \lambda \rho +  (1 - \lambda) \, (\tr \rho) \,
\ts{\frac{1}{d} }  I
\ee
where $I$ is the $d \times d$ identity matrix.   It is well-known (see e.g.,
\cite{Ki4}) that this map is CPT (CP and trace-preserving)
for values of $\lambda$ in the range
$ -{1 \over d^2 - 1} \leq \lambda \leq 1 $.
The map (\ref{def:dep}) is clearly EP for $0 \leq \lambda \leq 1$,
and hence Theorem \ref{thm2} can be used to show that
   $\nu_t \big (\Phi^{\ot m} \big) = \big[ \nu_t(\Phi) \big]^m$
for integer $t$.
   This result for products of depolarizing channels
   was first established in \cite{AH};
   subsequently, it was extended to all $t \geq 1$ in \cite{FH} and \cite{Ki4}.

In  \cite{GLR}   the map (\ref{def:dep}) was generalized by replacing
    ${\frac{1}{d} }  I$ by a fixed arbitrary density matrix   $\gamma$:
\be\label{def:dep-sh}
\rho \rightarrow \lambda \rho + (1 - \lambda) \,\, \tr \rho \,\, \gamma
\ee
Using a basis in which $\gamma$ is diagonal, it is easy to verify
that (\ref{def:dep-sh}) is CPT and EP  for $0 \leq \lambda \leq 1$.
Thus, Theorem \ref{thm2} can again be used to show that
   $\nu_t \big (\Phi^{\ot m} \big) = \big[ \nu_t(\Phi) \big]^m$
for all
integer $t$.   This result
   was established for $t = 2$ in \cite{GLR}.

\subsection{Positive Kraus operators}
If a channel has a Kraus representation $\Phi(\rho) = \sum A_k \rho A_{k}^{*}$
where each matrix $A_k$ is EP, then the map $\Phi$ is
EP, and   Theorem \ref{thm2} can be applied. In particular, this 
holds when  $A_k = \sqrt{p_k} P_k$ where $P_k$ is a permutation 
matrix, and $\sum
p_{k} = 1$. This is a particular case of the class of so-called
``random unitary''
channels.

\subsection{Maps which are not EP}
To give an example of a map which is not EP, it suffices to recall
that Werner and Holevo \cite{WH} found maps for which
(\ref{thm2:eqn1}) is false   for $t$ sufficiently large; therefore, 
these maps cannot be EP.

\medskip
We now show that there are also qubit maps which are not
EP  by observing that (\ref{def:EP}) implies that
  $\tr E_{ij} \Phi(E_{k \ell})$ is real for all $i,j,k, \ell$ .
Let $\{ a_{jk} \}_{j,k =0,1,2,3} $ be the matrix representing the qubit map
$\Phi$ in the basis consisting of $\{ I, \sigma_1, \sigma_2, \sigma_3 
\}$, i.e.,
the identity and the three Pauli matrices with the implicit convention
$\sigma_0 = I$, and let $E_{ij} = |i \kb j|$ in the standard basis for
${\Bbb C}_m$.
Then, e.g.,
\bee
4 \, \tr \, E_{12} \Phi(E_{11})     & =   ~ \tr \, (\sigma_1 + i 
\sigma_2) \Phi(I + \sigma_3)  ~ =
&   (a_{10} + a_{13})  \, +  \, i  (a_{20} + a_{23}) \\
4 \, \tr \, E_{12} \Phi(E_{22})  & =   ~ \tr \, (\sigma_1 + i 
\sigma_2) \Phi(I - \sigma_3)  ~ =
&  (a_{10} - a_{13})  \, +  \, i  (a_{20} - a_{23}) .
\eee
Therefore, the requirement that
${\rm Im}  \tr \, E_{12} \Phi(E_{11}) = {\rm Im}  \tr \, E_{12} 
\Phi(E_{22}) = 0$ implies
that  $(a_{20} \pm a_{23}) = 0$ which implies  $a_{20} = a_{23} = 0$.
Proceeding in this way, one can show that a necessary condition for
$\tr E_{ij} \Phi(E_{k \ell})$ to be real is that
\be  \label{real}
a_{j2} = a_{2k} = 0 \quad {\rm for} \quad j,k = 0,1,3
\ee
i.e., all $a_{jk}$ with $j = 2$ or $k = 2$ vanish unless $j = k$.
The map (\ref{qubit.canon}) corresponds to the choice $a_{00} = 
1,a_{0k} = 0,  a_{j0} = t_k$ and
$a_{jk} =  \lambda_j \delta_{jk}, ~j,k =1,2,3$.
This map does not satisfy the condition (\ref{real})
when $t_2 \neq 0$.   Now recall that a change of basis on
${\Bbb C}_m$
corresponds to a rotation on ${\Bbb R}_3$.
As explained in Appendix  B of \cite{KR1}, making a change of
basis on the domain and
range of ${\Bbb C}_m$, corresponds to
changing
\be
{\bf v}  \rightarrow O_1 {\bf v}, \quad
T \rightarrow  O_1 T O_2
\ee
where $O_1$, $O_2$ are $3 \times 3$ orthogonal matrices,
$T$ is the $3 \times 3$ matrix $a_{jk}$ with $j,k = 1,2,3$ and
${\bf v} = (a_{01}, a_{02}, a_{03})^{T} $.   Thus, for the map 
(\ref{qubit.canon}),
$T$ has elements  $\lambda_j \delta_{jk}$, ${\bf v} = (t_{1}, t_{2}, 
t_3)^{T} $.
When all $\lambda_j \neq 0$ are
distinct and all  $t_j \neq 0$, any $O_1$ which makes $t_2 = 0$
will make  either $a_{21} \neq 0$ or $a_{23} \neq 0$, violating
(\ref{real}).   In general
there is no choice of $O_1$, $O_2$ for which  (\ref{real}) holds.

One can similarly show that qubit maps of the
form (\ref{qubit.canon})
with the additional restrictions above
do not satisfy the weaker condition that $\wh{\Phi} \circ \Phi$ is EP.
Note that $\wh{\Phi} \circ \Phi$ is represented by the $4 \times 4$
matrix $B \equiv A^* A$
(indexed by $0,1,2,3$).    When $t_2 = 0$, all elements of $B$
are explicitly non-negative except for $b_{12} = b_{21} = \lambda_1^2 
- \lambda_2^2$.
This will
be negative when $|\lambda_1|  <  |\lambda_2|$, which suggests
that maps which do not satisfy
(\ref{qt20}) do not satisfy the condition that $\wh{\Phi} \circ \Phi$ 
is EP.

\section{Proofs of Theorems}
Our proofs will use the following consequence of  H\"older's
inequality: for any
matrices $B_1, B_2, \dots$ and integer $n$,
\be\label{Hold1}
| \tr (B_1 \, B_2 \, \dots \, B_{n} ) | \leq || B_1 ||_{n} \, ||
B_{2} ||_{n} \, \dots \, ||B_{n} ||_{n} ~.
\ee
Furthermore the definition of the $(p \rightarrow q)$ norm implies that for
any matrix $B$ and linear operator $L$,
\be\label{p->q}
|| L(B) ||_{q} \leq || L ||_{p \rightarrow q} \, || B ||_{p} ~.
\ee

\subsection{Proof of Theorem \ref{thm1}}
Let $A$ be any $nm \times nm$ matrix, then
\be
A = \sum_{ij} E_{ij} \ot A_{ij}
\ee
where $\{ A_{ij} \}$ are the
$m \times m$ blocks. Hence
\be\label{tensor1}
(K \ot L) A = \sum_{ij} K( E_{ij}) \ot L(A_{ij})
\ee
For any integer $t$,
\be\label{tr1}
\tr | (K \ot L) (A) |^{2t} = \tr \Bigg( (K \ot L) (A) \,\, [(K \ot L)
(A)]^{*} \Bigg)^t
\ee
Using the representation (\ref{tensor1}) in (\ref{tr1}) we get
\be\label{tr2}
\tr | (K \ot L) A |^{2t} = \sum \tr \Big( K( E_{i_1 j_1}) \, K(
E_{i_2 j_2})^{*} \, \dots \Big) \,
\tr \Big( L(A_{i_1 j_1}) \, L(A_{i_2 j_2})^{*}  \dots \Big)
\ee
Now we apply (\ref{Hold1}) with $n=2t$ and (\ref{p->q}) with $q=2t$: this gives
\be
\tr | (K \ot L) A |^{2t}    \leq   \bigg( || L ||_{p \rightarrow 2t}
\bigg)^{2t} \,\,
    \sum \bigg| \tr \Big( K( E_{i_1 j_1}) \, K( E_{i_2 j_2})^{*} \,
\dots \Big) \bigg| \,
|| A_{i_1 j_1} ||_{p}   \dots
\ee
The assumption that $K$ is EP implies that
\be
\tr \Big( K( E_{i_1 j_1}) \, K( E_{i_2 j_2})^{*} \, \dots \Big) \geq 0
\ee
for all indices $i_1, j_1,\dots$. It follows that
\be
\tr | (K \ot L) A |^{2t}    \leq
\bigg( || L ||_{p \rightarrow 2t} \bigg)^{2t} \,\, \tr | K( \alpha) |^{2t}
\ee
where $\alpha$ is the $n \times n$ matrix with entries
\be
\alpha_{ij} = || A_{ij} ||_{p} = || A_{ij}^{*} ||_{p}
\ee
Finally we use the following result of Bhatia and Kittaneh \cite{BK}:
for $1 \leq p \leq 2$
\be  \label{eq:BK}
\tr \alpha^2 = \sum_{ij} \, || A_{ij} ||_{p}^2 \leq || A ||_{p}^2
\ee
This implies
\be
|| (K \ot L) A  ||_{2t} \leq || L ||_{p \rightarrow 2t} \,\, || K
||_{2 \rightarrow 2t} \,\,
|| \alpha ||_{2} \leq
    || L ||_{p \rightarrow 2t} \,\, || K ||_{2 \rightarrow 2t} \,\,
|| A ||_p
\ee
which completes the proof that
\be
|| K \ot L ||_{p \rightarrow 2t} \leq || K ||_{2 \rightarrow 2t} \,\,
|| L ||_{p \rightarrow 2t}
\ee
At $p=2$, equality can be achieved using  the product of states
which maximize  $|| K ||_{p \rightarrow 2t} $  and $|| L ||_{p \rightarrow 2t}$.
\qed

\subsection{Proof of Theorem \ref{thm2}}
Let $A \geq 0$ and write
\be
A = \sum_{ij} E_{ij} \ot A_{ij}
\ee
Since $\Phi \ot \Omega$ is positivity preserving it follows that
\be\label{thm2.pf1}
\tr (\Phi \ot \Omega) (A)^t = \sum \tr \bigg(\Phi(E_{i_1 j_1}) \,
\Phi(E_{i_2 j_2}) \, \dots \bigg) \,
\tr \bigg( \Omega(A_{i_1 j_1}) \, \Omega(A_{i_2 j_2}) \, \dots \bigg)
\ee

Using H\"older's inequality again as in (\ref{Hold1}), and using the
fact that $\Phi$ is EP,  we deduce
\be\label{CP1}
\tr (\Phi \ot \Omega) (A)^t \leq
\sum \tr \bigg(\Phi(E_{i_1 j_1}) \, \Phi(E_{i_2 j_2}) \, \dots \bigg) \,
|| \Omega(A_{i_1 j_1}) ||_t \, || \Omega(A_{i_2 j_2}) ||_t \, \dots
\ee
Since $(I \ot \Omega)(A) \geq 0$ it follows that for all $i,j$
\be \label{2block}
\Omega(A_{ij}) = \Omega(A_{ii})^{1/2} \, R_{ij} \, \Omega(A_{jj})^{1/2}
\ee
where $R_{ij}$ is a  contraction, that is $|| R_{ij} ||_{\infty} \leq 1$. Hence
\be  \label{2ineq}
|| \Omega(A_{ij}) ||_t  \leq  ||\Omega(A_{ii})||_{t}^{1/2} \, ||
\Omega(A_{jj})||_{t}^{1/2}
\ee
Substituting into (\ref{CP1}) we deduce
\be
\tr (\Phi \ot \Omega) (A)^t  \leq   \tr \, \Phi(\beta)^t
\ee
where now $\beta$ is the $n \times n$ matrix with entries
\be
\beta_{ij} =  ||\Omega(A_{ii})||_{t}^{1/2} \, || \Omega(A_{jj})||_{t}^{1/2}
\ee
Since $\beta \geq 0$ we deduce
\be
\tr (\Phi \ot \Omega) (A)^t  & \leq & \nu_{t}(\Phi)^t \,\, \bigg( \tr
(\beta) \bigg)^t \nonumber \\
& = & \nu_{t}(\Phi)^t \,\, \bigg( \sum_{i=1}^n ||\Omega(A_{ii})||_{t}
\bigg)^{t} \nonumber \\
& \leq & \nu_{t}(\Phi)^t \,\, \nu_{t}(\Omega)^t \,\, \bigg( \tr A \bigg)^t
\ee
Taking the $t^{\rm th}$ root of both sides and taking the $\sup$ over
$A$ shows that
\be
\nu_{t}(\Phi \ot \Omega) \leq \nu_{t}(\Phi) \,\, \nu_{t}(\Omega)
\ee
as required.
Equality can be achieved using a product of states which maximize 
$\nu_{t}(\Phi)$ and $\nu_{t}(\Omega)$.
\qed

\subsection{Proof of Theorem~\ref{thm4} }    First, observe that 
(\ref{2block}) only involves the
$2 \times 2$
   submatrix $\pmatrix{\Omega(A_{ii}) & \Omega(A_{ij}) \cr
\Omega(A_{ij}^*) & \Omega(A_{jj}) }$; therefore,
   the inequality (\ref{2ineq}) holds whenever $\Omega$ is 2-positive.
  Next, proceed as in the proof of Theorem~\ref{thm1} up to (\ref{tr2}).
  Since $\Phi$  is EP, one can then conclude that a variant  of (\ref{CP1})
   holds with $\Phi(E_{i_1 j_1}) \, \Phi(E_{i_2 j_2}) $ replaced by
   $\Phi(E_{i_1 j_1}) \,  [\Phi(E_{i_2 j_2})]^* $ and $t$ replaced by 
the even integer  $2t$.
   When $\Omega$ is 2-positive, the remainder of the proof of   Theorem~\ref{thm2} 
goes
   through to yield    $\nu_{2t}(\Phi \ot \Omega) \leq \nu_{2t}(\Phi) 
\, \nu_{2t}(\Omega)$.

\subsection{Proof of Theorem~\ref {CP-EP-inf}}
The proof is a transcription of the proof of Theorem \ref{thm2},
with matrices replaced by integral kernels.
First note that a bipartite state $R$ on $L^2({\Bbb R}) \ot L^2({\Bbb R})$
is described by a kernel $K(x_1, y_1; x_2, y_2)$.
For fixed $x_1$ and $x_2$ we define the function
\be
T_{x_1, x_2}(y_1, y_2) = K(x_1, y_1; x_2, y_2)
\ee
Then by analogy with (\ref{CP1}) we have
\be\label{inf-dim-CP}
&& \hskip -0.6in \tr (\Phi \ot \Omega)(R)^t \\
& & \hskip -0.6in = \int \dots \int \, \prod_{i=1}^t
G(x_i, x_{i+1}; u_i, v_i) \,
\tr \, \bigg(\Omega(T_{u_1, v_1})  \, \dots \Omega(T_{u_t, v_t}) \bigg)
  \prod_{i=1}^t d x_i \, d u_i \, d v_i \\
\ee
where we use the labelling convention $t+1\equiv1$.
Applying H\"older's inequality 
(see, for example, Appendix B of \cite{Rus}) gives
\be\label{Hold-inf}
\Big| \tr \, \bigg(\Omega(T_{u_1, v_1})  \, \dots \Omega(T_{u_t, 
v_t}) \bigg) \Big|
\leq || \Omega(T_{u_1, v_1}) ||_t \, \dots \,
|| \Omega(T_{u_t, v_t}) ||_t
\ee
The analog of (\ref{2ineq}) is
\be\label{2ineq-inf}
|| \Omega(T_{u, v}) ||_t \leq || \Omega(T_{u, u}) ||_t^{1/2} \,\,
|| \Omega(T_{v, v}) ||_t^{1/2}
\ee
Using the pointwise positivity of $G$, and
substituting (\ref{Hold-inf}) and (\ref{2ineq-inf})
back into (\ref{inf-dim-CP}) gives
\be
\tr (\Phi \ot \Omega)(R)^t \leq
\tr \Phi(S)^t
\ee
where $S$ is the operator with kernel
\be
S(u, v) = || \Omega(T_{u, u}) ||_t^{1/2} \,\, || \Omega(T_{v, v}) ||_t^{1/2}
\ee
Using
\be
\tr (S) & = & \int || \Omega(T_{u, u}) ||_t \, d u \\
& \leq & \nu_{t}(\Omega) \, \int \tr \, (T_{u,u}) \, d u \\
& = & \nu_{t}(\Omega) \, \int \int K(u, v; u, v) \, d u \, d v
\ee
the rest of the argument follows as before.

\subsection{Proof of Theorem \ref{bos-ch}}
In terms of integral
kernels, the channel (\ref{def:bosonic}) acts by
\be
\Phi\, : \, K(x,y) \rightarrow \int \, h(x-y, \beta) \, K(x-\beta, 
y-\beta) \, d \beta
\ee
where
\be\label{def:h-inf}
h(x-y, \beta) = \int \, P(\alpha, \beta) \, e^{i \alpha (x-y)} \, d \alpha
\ee
This can be re-written in the form (\ref{def:inf}) with
\be\label{corres}
G(x,y; u,v) = \delta(u-x+y-v) \, h(u-v, y-v)
\ee
The condition (\ref{P-pos}) implies that
$h(a,b) \geq 0$ for all $a, b$. By using a sequence of positive approximations
$\delta_n$ for the
$\delta$-function in (\ref{corres}), we obtain a sequence of positive 
kernels $G_{n}$
for which   Theorem \ref{CP-EP-inf} can be applied. The result is 
then obtained in the limit
$n \rightarrow \infty$.

\section{Conclusion}
The additivity conjecture arose in quantum information theory in the
context of entropy-related
properties of completely positive trace-preserving (CPT) maps.
In the course of seeking a proof of the conjecture,  Amosov and
Holevo \cite{AH}
proposed a more general multiplicativity result involving Schatten $p$-norms.
In this larger context it is natural to drop the trace-preserving
condition, and consider just
completely positive maps, and most of the known results hold for this
more general class.
A further natural generalization of the question is to consider
multiplicativity properties involving $p$ to $q$ norms of CP maps for general
values $p,q \geq 1$. In this paper we have
demonstrated some multiplicativity results in this case for  large
classes of maps characterized by  conditions which are not
equivalent to the CP property in the case of non-commutative systems. 
This may be an indication that the multiplicativity property
has its roots in a different setting.

\bigskip
\noindent {\bf Acknowledgement:}
The work of CK and MN was supported in part by the
National Science Foundation under grants DMS--0101205 and DMS--0400426;
that of MN and MBR was supported in part    by  the National Security 
Agency (NSA) and Advanced Research and Development Activity (ARDA) 
under Army Research Office (ARO) contract number  DAAD19-02-1-0065; 
and that of MBR by the National Science Foundation under Grant 
DMS-0314228.

 {~~}

\end{document}